\documentstyle[preprint,tighten,aps,amssymb,psfig]{revtex}
\begin{document}

\draft

\title{Unified description of light- and strange-baryon spectra}

\author{L. Ya. Glozman,${}^1$ W. Plessas,${}^1$
K. Varga,${}^{2,3}$ and R. F. Wagenbrunn${}^1$ }
\address{${}^1$Institute for Theoretical Physics, University of Graz,
A-8010 Graz, Austria\protect\\
${}^2$Institute of Nuclear Research, Hungarian Academy of Sciences,
H-4001 Debrecen, Hungary\protect\\
${}^3$RIKEN, Hirosawa 2-1, Wako, Saitama 35101, Japan}

\date{\today}
\maketitle
\begin{abstract}
We present a chiral constituent quark model for light and strange baryons
providing a unified description of their ground states and excitation spectra. 
The
model relies on constituent quarks and Goldstone bosons arising as effective
degrees of freedom of low-energy QCD from the spontaneous breaking of chiral
symmetry. The spectra of the three-quark systems are obtained from a precise
variational solution of the Schr\"odinger equation with a semirelativistic
Hamiltonian. The theoretical predictions are found in close agreement with
experiment.
\end{abstract}
\pacs{12.39.-x, 14.20.-c, 21.45.+v}

%\narrowtext

An intricate question of low-energy quantum chromodynamics (QCD) is the one
after the effective degrees of freedom that govern the physics of light and
strange baryons. The early (naive) quark model \cite{kokk} was successful in
classifying hadrons and describing some gross properties of their spectra but 
no firm
evidence in the dynamics of the valence quarks was achieved. Even when,
motivated by QCD, the concept of one-gluon exchange (OGE) \cite{rujula} was
introduced as interaction between confined constituent quarks, a number of
delicate problems remained unsolved. In this context up till now one has been
unable to explain, e.g., the correct level orderings in light- and strange
baryon spectra \cite{gloris,gloschl}, the spin content of the nucleon
\cite{ashman}, or the flavor asymmetry of the sea in the nucleon
\cite{gottfr,dryan}. The shortcomings essentially stem from the fact that the
implications of spontaneous breaking of chiral symmetry (SB$\chi$S) are not
properly taken into account in such a model, and as a consequence the pertinent
interactions between constituent quarks turn out inadequate. Evidently, if one
assumes constituent quarks of flavors $u$, $d$, $s$ with masses considerably
larger than the corresponding current-quark masses, this already means that the
underlying chiral symmetry of QCD is spontaneously broken. As a consequence of
that SB$\chi$S, at the same time Goldstone bosons appear, which couple
directly to the constituent quarks \cite{weinberg,mangeo,diakonov}. Hence,
beyond the scale of SB$\chi$S one is left with constituent quarks with
dynamical masses related to $<\bar{q}q>$ condensates and with Goldstone bosons 
as the
effective degrees of freedom. This feature that in the Nambu-Goldstone mode of
chiral symmetry, constituent-quark and Goldstone-boson fields prevail
together is also well supported, e.g., by the $\sigma$-model \cite{gell2} or
the Nambu--Jona-Lasinio model \cite{nambu}. In the same framework also the
problems with the spin and flavor content of the nucleon get naturally
resolved \cite{chengli}. As a consequence, baryons are to be considered as
systems of three constituent quarks that interact by Goldstone-boson exchange
(GBE) and are subject to confinement \cite{gloris,gloschl}.

The Goldstone bosons manifest themselves in the octet of pseudoscalar mesons
($\pi$, $K$, $\eta$). In the large-$N_C$ limit, when the axial anomaly vanishes
\cite{witten}, the spontaneous breaking of the chiral symmetry
$U(3)_L\times U(3)_R\to U(3)_V$ implies a ninth Goldstone boson \cite{coleman},
which corresponds to the flavor singlet $\eta'$. Under real conditions,
for $N_C=3$, a certain contribution from the flavor singlet remains and the
$\eta'$ must thus be included in the GBE interaction.

In view of these considerations we propose a semirelativistic
chiral constituent quark model that is based on the following three-quark
Hamiltonian
\begin{equation}
\label{htot}
H=\sum_{i=1}^3\sqrt{\vec{p}_i{}^2+m_i^2}+\sum_{i<j=1}^3V_{ij}.
\end{equation}
Here the relativistic form of the kinetic-energy operator is employed, with
$\vec{p}_i$ the 3-momentum and $m_i$ the masses of the constituent quarks. The
dynamical part consists of the quark-quark interaction
\begin{equation}
V_{ij}=V_{\chi}+V_{conf},
\end{equation}
where the confinement potential is taken in the linear form
\begin{equation}
V_{conf}(r_{ij})=V_0+C r_{ij},
\end{equation}
and the chiral potential $V_\chi$ is derived from GBE. Of the latter it is
the spin-spin component that provides by far the most dominant contribution in
baryons; it is manifested by the sum of octet and singlet pseudoscalar
meson-exchange potentials \cite{gloris,gloschl}
\begin{eqnarray}
\label{voct}
V_\chi^{octet}(\vec r_{ij})  &=&
\left[\sum_{F=1}^3 V_{\pi}(\vec r_{ij}) \lambda_i^F \lambda_j^F\right.
+\sum_{F=4}^7 V_K(\vec r_{ij}) \lambda_i^F \lambda_j^F
\nonumber \\ 
&&\left.\raisebox{0ex}[3ex][3ex]{}+V_{\eta}(\vec r_{ij}) \lambda_i^8 
\lambda_j^8\right]
\vec\sigma_i\cdot\vec\sigma_j,
\end{eqnarray}
\begin{equation}
\label{vsing}
V_\chi^{singlet}(\vec r_{ij}) = \frac {2}{3}\vec\sigma_i\cdot\vec\sigma_j 
V_{\eta'}(\vec r_{ij}),
\end{equation}
where  $\vec{\sigma_i}$ and $\lambda_i^F$ represent the quark spin and
flavor matrices, respectively. In the simplest derivation, when pseudoscalar
or pseudovector couplings are employed at point-like meson-quark vertices
and the boson fields satisfy the linear Klein-Gordon equation, one obtains, in
static approximation, the well-known meson-exchange potentials
\begin{equation}
\label{vgamma}
V_\gamma (\vec r_{ij})= \frac{g_\gamma^2}{4\pi}%\frac{1}{3}
\frac{1}{12m_im_j}
\left\{\mu_\gamma^2\frac{e^{-\mu_\gamma r_{ij}}}{ r_{ij}}-
4\pi\delta (\vec r_{ij})\right\},
\end{equation}
with $\mu_\gamma$ ($\gamma=\pi,K,\eta,\eta'$) being the individual 
phenomenological meson masses, and
$g_\gamma^2/4\pi$ the meson-quark coupling constants. In general,
the structure of the quark-quark potential in momentum-space representation
is
\begin{equation}
V_\gamma(\vec{q})\sim\vec{\sigma_i}\cdot\vec{q}\;\vec{\sigma_j}\cdot\vec{q}
D(q^2)F^2(q^2),
\end{equation}
where $D(q^2)$ is the dressed Green function for the chiral field,
including both nonlinear terms of the chiral Lagrangian and fermion loops, and
$F(q^2)$ is a meson-quark form factor, which takes into account the extended
structure of the quasiparticles. In the limit $\vec{q}\to 0$, one has
$D(q^2)\to -(\vec{q}\;^2+\mu^2)^{-1}\ne\infty$ and $F(q^2)\to 1$, and
consequently $V_\gamma(\vec{q}=0)=0$.
Therefore the pseudoscalar meson-exchange interaction has to fulfill the 
requirement of
the volume integral to vanish: $\int d^3rV_\gamma(\vec{r})=0$. 
Since at large distances 
$V_\gamma\sim \mu_\gamma^2\frac{e^{-\mu_\gamma r}}{r}$,
there must be a strong short-range part of opposite sign in order to guarantee
the volume integral constraint. Inside baryons this short-range part dominates
over the Yukawa tail and it becomes of crucial importance to reproduce the
baryon spectra.

A suitable parametrization of the GBE potential, preserving a zero volume
integral, is thus given by
\begin{equation}
\label{vyuk}
V_\gamma (\vec r_{ij})= \frac{g_\gamma^2}{4\pi}%\frac{1}{3}
\frac{1}{12m_im_j}
\left\{\mu_\gamma^2\frac{e^{-\mu_\gamma r_{ij}}}{ r_{ij}}-
\Lambda_\gamma^2\frac{e^{-\Lambda_\gamma r_{ij}}}{ r_{ij}}\right\}.
\end{equation}
It  involves the parameters $\Lambda_\gamma$ corresponding to the
individual exchanged mesons. Clearly the values of $\Lambda_\gamma$ should vary
with the magnitudes of the meson masses $\mu_\gamma$: with a larger meson mass
$\mu_\gamma$ also $\Lambda_\gamma$ should become larger. Otherwise the
individual meson-exchange potentials in (\ref{vyuk}) could receive
unwarranted contributions (e.g., a certain meson-exchange contribution could
become attractive instead of repulsive or vice versa at short distances). 
In order to avoid a
proliferation of free parameters, by assuming 4 independent values of
$\Lambda_\gamma$ (for each $\gamma=\pi,K,\eta,\eta'$), we adopt the linear 
scaling
prescription
\begin{equation}
\label{linear}
\Lambda_\gamma=\Lambda_0+\kappa\mu_\gamma
\end{equation}
which involves only the 2 free parameters $\Lambda_0$ and $\kappa$.

Due to the explicit chiral symmetry breaking in QCD the various quark-meson 
coupling
constants could naturally be different. Again, we try to keep the number of free
parameters as small as possible and assume a single octet-quark coupling
$g_8^2/4\pi$ for all octet mesons ($\pi$, $K$, $\eta$). Its value can be
extracted from the phenomenological pion-nucleon coupling constant as 
$g_8^2/4\pi=0.67$
\cite{gloris}.
Because of the particular character of the $\eta'$ meson (cf., the discussion
above), the flavor-singlet coupling constant may well be different from
the octet one. This assumption is also supported by the successful explanation
of the flavor and spin content of the nucleon \cite{chengli}. 
Therefore we treat the ratio
$(g_0/g_8)^2$ as a free parameter.

For the constituent quark masses we take the typical values $m_u=340$ MeV and
$m_s=500$ MeV.
Considering the constituent-quark as well as meson masses and the octet
coupling constant as predetermined, the GBE potentials of Eq.
(\ref{vyuk}) involve only 3 free parameters. Their values,
together with the 2 free parameters of the confinement potential, 
were determined
from a fit to the baryon spectra. The resulting numerical values are given in
Table \ref{potpar} together with the fixed model parameters.
We remark that the parameters given in Table \ref{potpar} are only one choice 
out
of a possible set of others that lead to a similar quality of description of
the baryon spectra. In case the fixed parameters were chosen differently,
e.g., with regard to the specific values of the constituent quark masses or the
octet coupling constant, the free parameters would get slightly changed but a
similar fit could be achieved. The baryon spectra alone do simply not guarantee
for a unique determination of the model parameters. Further studies of other
observables are
necessary to constrain their values.
Nevertheless, it is pleasing to find the present parameter values of
reasonable magnitudes. For example, the confinement strength is comparable with
the string tension extracted from lattice calculations \cite{creutz} and it is
also consistent with the slopes of Regge trajectories.

The three-quark system with the Hamiltonian of Eq. (\ref{htot}) is treated by
solving the Schr\"odinger equation with the stochastic variational method
\cite{varga}. This technique has been tested in a number of benchmark cases
before. The results prove reliable to an accuracy of better than $1\%$ in the
present calculation.
In Fig. 1 we show the predictions of our model for all
light- and strange-baryon excitation levels up to $M\lesssim 1850$ MeV; the
nucleon is normalized to its mass of 939 MeV (what determines the value of
the confinement potential parameter $V_0$). All masses corresponding to
three- and four-star resonances in the most recent compilation of the PDG
\cite{pdg} are included.

From the results it is immediately evident that quite a satisfactory
description of the spectra of all low-lying light and strange baryons is
achieved in a unified framework. In particular, the
level orderings of the lowest positive- and negative-parity states in the
nucleon spectrum are reproduced correctly, with the $\frac{1}{2}^+$
Roper resonance $N(1440)$ falling well below the negative-parity
$\frac{1}{2}^-$ and $\frac{3}{2}^-$ states $N(1535)$ and $N(1520)$, 
respectively.

Likewise, in the $\Lambda$ and $\Sigma$ spectra the positive-parity
$\frac{1}{2}^+$ excitations $\Lambda(1600)$ and 
$\Sigma(1660)$ fall below the negative-parity
$\frac{1}{2}^-$-$\frac{3}{2}^-$ states 
$\Lambda(1670)$-$\Lambda(1690)$ and the
$\frac{1}{2}^-$ state $\Sigma(1750)$, respectively.
In the $\Lambda$ spectrum,
at the same time, the negative-parity $\frac{1}{2}^-$-$\frac{3}{2}^-$ states
$\Lambda(1405)$-$\Lambda(1520)$ remain the
lowest excitations above the $\Lambda$ ground state. By 
the correct level orderings of the positive- and negative-parity states a
long-standing problem of baryon spectroscopy is resolved.
At this stage, only one state is not reproduced in agreement or close to
experiment, the flavor singlet $\Lambda(1405)$ (we mention a possible reason
below).

The remarkable successes of the GBE quark-quark interaction of Eqs.
(\ref{voct}), (\ref{vsing}), and (\ref{vyuk}) is, of course, brought about 
by the particular symmetry
introduced through the
spin-flavor operators 
$\vec\sigma_i\cdot\vec\sigma_j\vec\lambda_i^F\cdot\vec\lambda_j^F$ and by the
short-range part of the interaction with a proper sign \cite{gloris,gloschl}. 
This makes the GBE potential just adequate
for the level structures found in experiment, and thus a unified description of
all light- and strange-baryon spectra is possible, even though
our model in the present
simplest version involves only a handful of free parameters. The action of the
chiral potential $V_\chi$ on the energy levels becomes especially transparent
when the coupling constant is gradually increased, see Fig. 2. Starting out
from the case with confinement only, one observes that with increasing coupling
the inversion of the 
lowest positive- and negative-parity states $N(1440)$ and
$N(1535)$-$N(1520)$ in the $N$ spectrum is achieved. At the same time
the level crossing of the corresponding states $\Lambda(1600)$ and
$\Lambda(1405)$-$\Lambda(1520)$
in the $\Lambda$ spectrum is avoided, just as demanded by phenomenology.

At this instance, a remark is in order about the necessity of employing a
relativistic kinetic-energy operator in the 3-quark Hamiltonian (\ref{htot}).
Certainly, this is only an intermediate step towards a fully covariant
treatment but it already allows to include the kinematical relativistic
effects. In any nonrelativistic approach these effects get compensated into the
potential parameters, which will not only assume unrealistic values
(cf., e.g., our previous nonrelativistic model \cite{gpp}) but one is also
faced with such disturbing consequences as $v/c>1$ (where $v$ is the mean 
velocity
of the constituent quark and $c$ is the velocity of light).

At the present stage, tensor forces are not yet included in our model. 
However, we
have already made estimates and numerical tests of their influences. They turn
out to be much less important for baryon masses, as compared to the spin-spin
part, at least for
the states considered in Fig. 1. 
It is clear also from phenomenology that
tensor forces can play only a subordinate role as the splittings of
corresponding $LS$-multiplets are generally small.

So far, the constituent quark model derived from GBE provides a reasonable 
description of
the light- and strange-baryon spectra. Nevertheless it needs further
improvements in many respects. 
For example, the coupling to decay channels should be explicitly included by
providing in addition to $QQQ$ further Fock components such as $QQQ\pi$,
$QQQK$, $QQQ\eta$, and $QQQ\eta'$. This will affect especially those states
lying close to continuum thresholds. One may expect, in particular, that
thereby the $\Lambda(1405)$ level will be shifted down since it lies close to
$\bar{K}N$ threshold \cite{lambda}. Furthermore, the
GBE quark-quark interaction should be tested with regard to observables other 
than
the spectra in order to obtain additional constraints.
It will be
interesting to see how far the description of light and strange baryons in
terms of constituent quarks and Goldstone bosons as effective degrees of
freedom can be driven.

The authors acknowledge valuable discussions with D. O. Riska and M. Rosina on
several aspects of the GBE interaction. The work was supported by the
Paul-Urban foundation and by OTKA grant No. T17298.

\begin{figure}[t]
\psfig{file=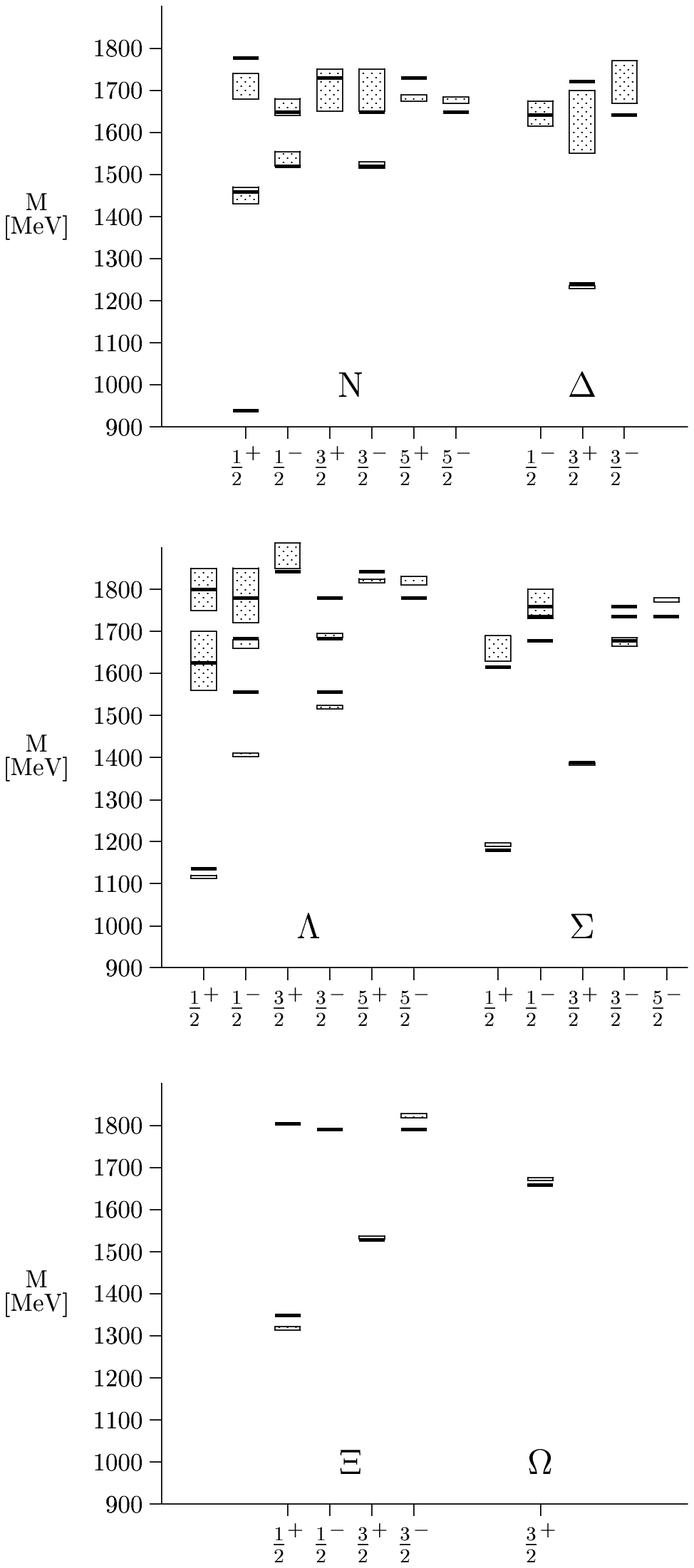,height=18cm}
\caption{Energy levels of the lowest light- and strange-baryon states
with total angular momentum and parity $J^P$. The nucleon ground
state is 939 MeV. The shadowed boxes
represent the experimental values with their uncertainties. The $\Delta$,
$\Sigma^\ast$, and $\Xi^\ast$ ground-state levels practically fall into their
rather tight experimental boxes.}
\end{figure}

\begin{figure}[tb]
\psfig{file=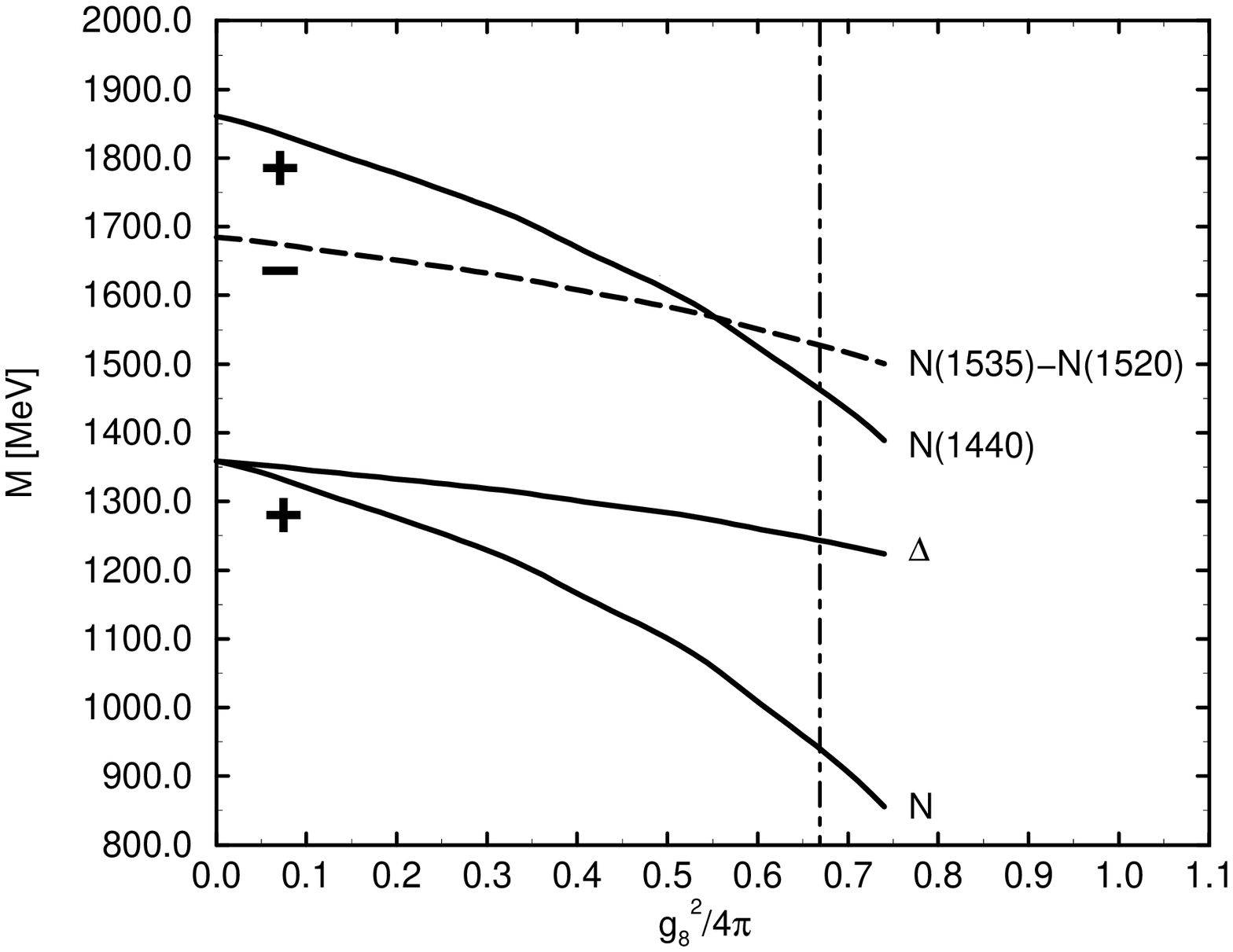,height=8cm}
\psfig{file=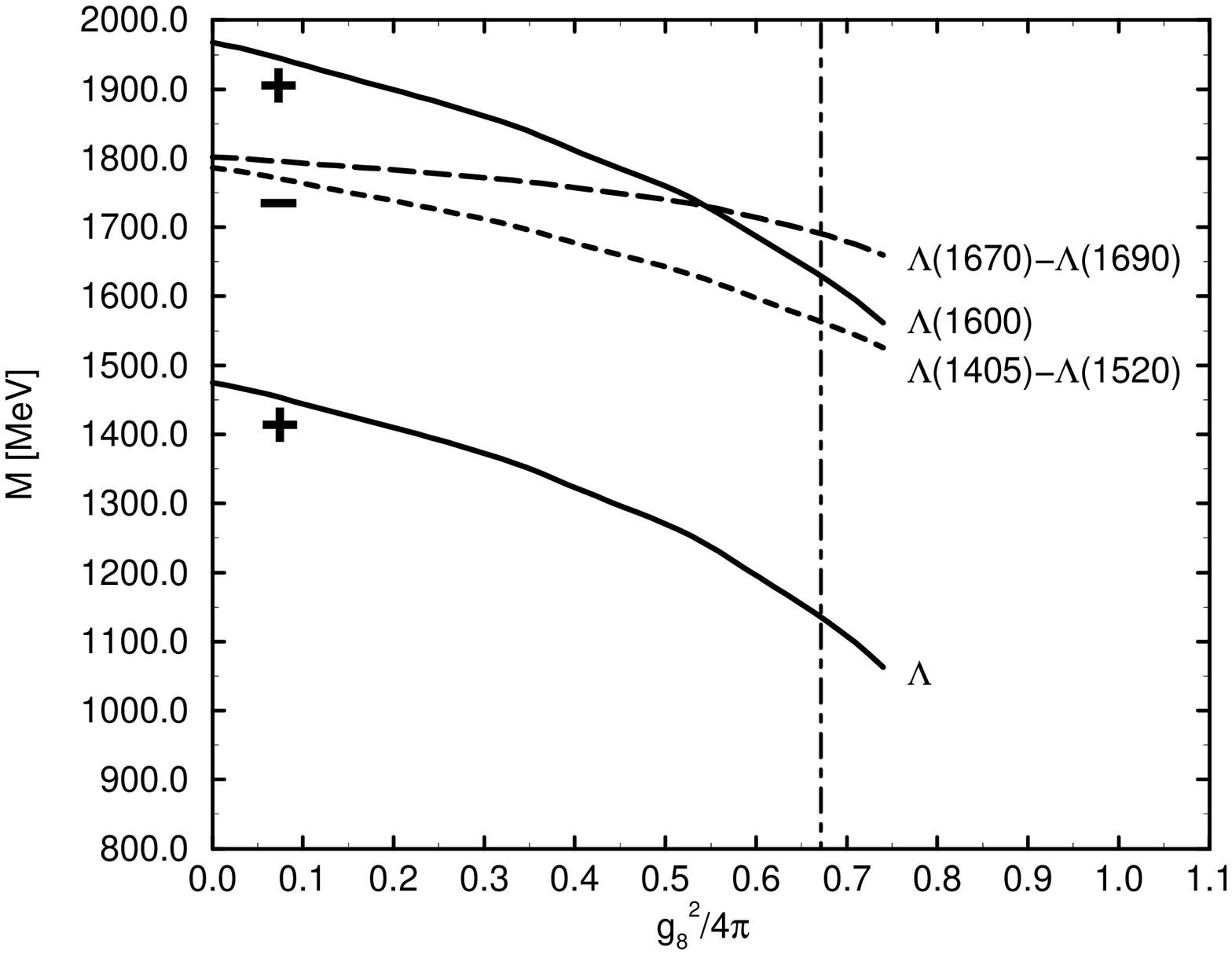,height=8cm}
\caption{Level shifts as a function of the strength of
the Goldstone-boson exchange interaction. Solid and dashed lines correspond to 
positive- and
negative-parity states, respectively.}
\end{figure}

\begin{table}[tb]
\caption{Parameters of the semirelativistic constituent quark model based on
GBE.}
\label{potpar}
\renewcommand{\arraystretch}{1.3}
\begin{tabular}{cccccccccc}
\multicolumn{10}{c}{Fixed parameters}\\ \hline
\multicolumn{3}{c}{Quark masses $[$MeV$]$}&&
\multicolumn{4}{c}{Meson masses $[$MeV$]$}&&\\
%\multicolumn{3}{c}{masses in MeV}&&
%\multicolumn{4}{c}{in MeV}&&
%$\frac{g_8^2}{4\pi}$\\
\multicolumn{2}{c}{$m_u$, $m_d$}&$m_s$&&$\mu_\pi$&$\mu_K$&$\mu_\eta$&
%$m_u$&$m_d$&$m_s$&&$\mu_\pi$&$\mu_K$&$\mu_\eta$&
$\mu_{\eta'}$&&$\frac{g_8^2}{4\pi}$\\ \hline
\multicolumn{2}{c}{340}&
%\parbox{0.8cm}{\centering 340}&
%\parbox{0.8cm}{\centering 340}&
\parbox{0.8cm}{\centering 500}&
&
\parbox{0.8cm}{\centering 139}&
\parbox{0.8cm}{\centering 494}&
\parbox{0.8cm}{\centering 547}&
\parbox{0.8cm}{\centering 958}&&
\parbox{0.8cm}{\centering 0.67}\\ \hline\hline
%\multicolumn{3}{c}{\parbox{2cm}{\centering $m_u=m_d=340$ MeV\\$m_s=500$ MeV}}&
%\multicolumn{2}{c}{\parbox{2cm}{\centering $\mu_\pi=139$ MeV\\$\mu_K=498$ MeV}}&
%\multicolumn{2}{c}{\parbox{2cm}{\centering $\mu_\eta=547$
%MeV\\$\mu_{\eta'}=958$ MeV}}& 
%\multicolumn{3}{c}{$\frac{g_8^2}{4\pi}=0.67$}\\ \hline\hline
\multicolumn{10}{c}{Free parameters}\\ \hline
\multicolumn{2}{c}{$(g_0/g_8)^2$}&
\multicolumn{2}{c}{$\Lambda_0$ $[\rm fm^{-1}]$}&
\multicolumn{2}{c}{$\kappa$}&
\multicolumn{2}{c}{$V_0$ $\rm [MeV]$}&
\multicolumn{2}{c}{$C$ $[\rm fm^{-2}]$}\\ \hline
\multicolumn{2}{c}{1.34}&
\multicolumn{2}{c}{2.87}&
\multicolumn{2}{c}{0.81}&
\multicolumn{2}{c}{-416}&
\multicolumn{2}{c}{2.33}
\end{tabular}

\end{table}

\end{document}